\documentclass{llncs}
\usepackage{makeidx,graphicx}  
\usepackage{epsfig}
\usepackage{wrapfig}
\usepackage{subfigure}

\begin{document}

\mainmatter              

\title{A Beginner's Guide to Systems Simulation in Immunology}

\author{Grazziela P. Figueredo \and Peer-Olaf Siebers  \and Uwe Aickelin \and Stephanie Foan}

\authorrunning{G. Figueredo et al.}   
%
\institute{Intelligent Modelling and Analysis Research Group, School of Computer Science, \\ The University of Nottingham, NG8 1BB, UK \\ gzf, uxa, pos, sjf@cs.nott.ac.uk}



\maketitle

\begin{abstract}
\noindent

Some common systems modelling and simulation approaches for immune problems are Monte Carlo simulations, system dynamics, discrete-event simulation and agent-based simulation. These methods, however, are still not widely adopted in immunology research. In addition, to our knowledge, there is few research on the processes for the development of simulation models for the immune system. Hence, for this work, we have two contributions to knowledge. The first one is to show the importance of systems simulation to help immunological research and to draw the attention of simulation developers to this research field. The second contribution is the introduction of a quick guide containing the main steps for modelling and simulation in immunology, together with challenges that occur during the model development. Further, this paper introduces an example of a simulation problem, where we test our guidelines.

\end{abstract}

\section{Introduction}
\label{Introduction}

Some important advances in immunology were facilitated by the joint work of immunologists and mathematicians~\cite{Louzoun:2007}. Many concepts existing in theoretical immunology are the result of mathematical models. Inumerous existing models in immunology are based on sets of ordinary differential equations (ODEs)~\cite{Gruber:2003,Eftimie2010}. This approach for immunology, however, in practice limits the modelling effort to simpler dynamics involving fewer immune elements such as cells or molecules and it only allows analysis at an aggregate level.  Moreover, it is not trivial to model problems involving individual localisation, memory of past events (or states) and emerging properties mathematically~\cite{Louzoun:2007}. Hence, systems simulation emerged as a complement of mathematics that allows to overcome some of these limitations. Moreover, systems simulation modelling methods are closer to the natural description of the system, without the need of an in depth understanding of mathematics~\cite{Bonabeau:2002}.

Sauro {\it et al.}~\cite{Sauro:2006} debates the usefulness of simulation in contrast to reductionism in biology. In reductionism, the dynamics of  complex systems can be understood from studying the properties of their parts~\cite{Conscilience:2007}. In contrast to reductionism, in holism (in its methodological version) the properties of the parts contribute to our understanding of the whole, but the properties can only be fully understood through the dynamics of the whole~\cite{Conscilience:2007}. Systems simulation is, therefore, based on holism. For biology, Sauro {\it et al.}~\cite{Sauro:2006} state that ``{\it reductionism has proven to be a highly successful strategy and has enabled us to uncover the molecular details of biological systems in unprecedented detail}''. The success of reductionist methods raised some scepticism as to the need for alternative approaches, such as systems biology. The challenge for simulation is, therefore, to generate novel insights that cannot be uncovered just by looking at a phenomena using reductionism. Examples of successful simulation approaches that helped advance immunological research were introduced in~\cite{Nuno:2007}. The models reviewed simulate interactions of immune cells and chemical substances, humoral responses and drug testing. With these simulations it is possible to observe emergent behaviour in the systems, which is not feasible with reductionism.

As a first objective of this study, therefore, we want to show that there is a distinct place for simulation in the tool set that aids advances in immunology. Moreover we want to show that there is a wide range of problems in immunology to be tackled by computer scientists and simulation developers.

As there few examples on the methodology for constructing immune systems simulations~\cite{Kitano:2002:Nature,Kitano:2002:Science,CoSMoS-2010-01-Y4}, the second objective is to introduce general guidances for conducting simulation studies in immunology and outline the challenges that might be encountered during the development of a simulation model. These guidances adapted from the work developed by Robinson~\cite{Robinson:2004} for operational research, considering the characteristics observed in immune simulations. We complement the current methodologies for constructing immune systems simulations by presenting a framework containing a life-cycle with the main steps to be followed by any developer, independent of the simulation modelling method chosen.

The remainder of the paper is organized as follows. In Section~\ref{SimulationMethods} we review the main characteristics of the simulation methods used in immunology. In Section~\ref{steps} we present the main steps and propose a life-cycle for conducting a simulation study for immune problems. In Section~\ref{CaseStudy} we present a case study where we develop a system dynamics simulation model from the steps outlined. In Section~\ref{Conclusions}, we present our conclusions and future work.

\section{Simulation Methods}
\label{SimulationMethods}

The choice of a modelling technique for a problem is driven by the resources available such as experimental data, an understanding of the mechanisms involved, the hypothesis to be tested and the level of abstraction needed to test the hypothesis. Once the conceptual model is defined, a simulation method needs to be chosen. There is a wide spectrum of simulation methods used in immunology. These methods are classified as static or dynamic, stochastic or deterministic, continuous or discrete. These methods model the system using either top-down or bottom-up perspectives. Static models help understand connections between system components, without explicitly representing time~\cite{Silva:2010}. Dynamic models aid in understanding dynamic implications and consequences over time of a system structure~\cite{Babulak:2010}. Deterministic models do not contain any probabilistic components. Stochastic models, on the other hand, consider random components~\cite{Banks:2005}. The characteristic of being continuous or discrete determine whether time and variables of the model change continuously or discretely. For example the age of an individual changes continuously in time, whereas the number of immune cells that die with age is a discrete value. Top-down approaches focus on the system at an aggregate level, while bottom-up approaches split the system in individual parts that will interact giving rise to the behaviour of the system as a whole~\cite{Macal:2010}.

Sauro {\it et al.}~\cite{Sauro:2006} mentions that the construction of models of biochemical and cellular behaviour has been traditionally carried out through a bottom-up approach, which combines laboratory data and knowledge of a reaction network to produce a dynamic model. This process, however, requires the reaction network to be known and the possibility to carry out the various laboratory experiments. Furthermore, the modelling relies on the fact that data from laboratory experimentation matches real-world phenomena, which is not always correct. Samples can be compromised during collection or during the experimentation process. In addition, although bottom-up approaches are very useful for immunology, there are circumstances where they can not be applied. Examples include when the reaction network process is not well understood, or laboratory experiments are known not to be able to reproduce the real-world reactions (for instance, given the environmental differences such as temperature). In addition, the authors argue that ``{\it top-down modelling strategies are closer to the spirit of systems biology exactly because they make use of systems-level data, rather than having originated from a more reductionist approach of molecular purification}''. The conclusion reached in their study was that there is no best approach as it is preferable to view them as complementary. Their ideas match other studies in biology and other research areas, which investigate the merits of each approach and their combination for simulation~\cite{Schieritz:2003}. To our knowledge, the most common system simulation approaches for immunology are Monte Carlo simulation, system dynamics, discrete-event simulation, cellular automata and agent-based simulation.

Monte Carlo simulations~\cite{Metropolis:1949} are largely used in molecular theoretical immunology. These techniques generate random numbers and observe that fraction of the numbers that obey a certain property (of properties). These methods are suitable for obtaining numerical solutions for problems as an alternative to their analytical solution. The disadvantage of these methods are that they do not provide information of how the elements of the system change during the simulation (dynamics of the system). Instead, they focus on the determination of the system outcome given a certain input, not taking time into account.

System dynamics (SD) is a top-down modelling technique that uses stocks, flows and feedback loops as concepts to model the behaviour of complex systems in a stock and flow diagram. It is an aspect of systems theory that is initially applied in order to understand complex aggregate behaviours in industry~\cite{Forrester:1961}. Currently, SD is applied to any complex system characterized by interdependency, mutual interaction, information feedback and circular causality. System dynamics simulation (SDS) is a continuous simulation for an SD model. It consists of a set of ODEs that are solved for a certain time interval.  These ODEs, however, are implicit in the system's structure and the relationships between the elements modelled can be established with experimental data. Some examples of SD applied to the immune system are found in ~\cite{Figueredo:2011,Foan:2011}.

Discrete-event simulation is also a top-down approach that models a system as a set of entities being processed and evolving over time according to the availability of resources and the triggering of events. The simulator maintains an ordered queue of events~\cite{Siebers:2007}. Each event occurs at an instant in time and marks a change of state in the system. It is process-oriented and the entities involved are passive~\cite{Robinson:2004}, with no pro-activity. The entities are individually represented and can be tracked throughout the system simulation. The models are stochastic and outputs usually represent average values~\cite{Robinson:2009}. Examples of applications for the immune system are found in~\cite{Look:1981,Zand:2004,Figge:2005}.

Cellular automata is a discrete model consisting of two main components. The first component is an infinite regular grid of cells, which constitutes the universe or space of the cellular automata. In computer simulations, however, due to space limitations, the cellular automata space is predetermined and finite. The second component is a finite automaton (or cell). Each cell from the grid contains a finite number of states and a predefined set of cells called {\it neighbourhood}. The communication of a cell with other cells within its neighbourhood is local, deterministic, uniform and synchronous~\cite{WOLFRAM:1983}. Each cell is initialized with an initial state at time $t = 0$. As time advances, the cells are updated according to a fixed rule, which is, in general, a mathematical function. This rule defines the next state of each cell according to its current state and its neighbourhood states. Examples of several applications for the immune system are found in~\cite{Nuno:2007}.

Agent-based simulation is a technique that employs autonomous agents that interact with each other~\cite{Macal:2005}. The agents' behaviour is described by rules that determine how they learn, interact with each other and adapt. The overall system behaviour arises from the agents' individual dynamics and their interactions~\cite{Siebers:2007}. For immunology, it can amalgamate {\it in vitro} data on individual interactions between cells and molecules of the immune system to build an impression of the system as a whole. Cellular automata and agent-based simulation have some similarities such as individual rules and interactions between the individuals entities. Moreover, both are bottom-up approaches capable of representing emergent behaviour in the system. The entities in cellular automata, however, do not have memory and only interact with individuals from the predefined neighbourhood, as their location does not change. In agent-based simulation, on the other hand, the agents are individual entities with memory and are capable to interact with any other agent in the system. Several examples of agent-based models in immunology are found in the european virtual human immune system project (ImmunoGrid)~\cite{ImmunoGrid}.

By reviewing the simulation methods and some of their applications to immune problems, such as those from~\cite{Nuno:2007,Look:1981,Zand:2004,Figge:2005,Thorne:2007}, it is possible to outline the benefits of simulation to immunology and, therefore, achieve our first objective of this work. Compared with real-world experimentation, simulation is time and cost-effective. Most laboratory experiments are expensive and have to be in agreement with ethical specifications. Furthermore, in a simulation environment, it is possible to systematically generate different scenarios, conduct and replicate experiments.

In the following section we fulfil our second research goal of this paper, which is to introduce a descriptive guide for the development of a simulation model in immunology and the challenges that might be encountered during this process. These guidances are kept general and can be applied to all simulation methods in immunology.

\section{Steps in a Simulation Study}
\label{steps}

As there is not much done to established guidance to develop simulations for the immune system, we studied those developed by~\cite{Robinson:2004} for simulation in operational research problems and adapted them for simulations of the immune system. The adaptation was performed by studying several simulations under different approaches developed for the immune system, as mentioned in the previous section. In order to adapt the guidelines developed by~\cite{Robinson:2004}, we  observed the similarities and differences with operational research and outlined general steps for building immune simulations. These steps represent a life-cycle of a simulation and therefore the method we present is iterative. Furthermore, we discuss the pitfalls that might be encountered during the process, as shown below. In some of the steps, extra efforts specific to immunology were not added as we believe they are generic for any type of simulation.

{\bf 1. Define the Objectives.} Overall, the objectives are either to investigate a theory or propose a ``what-if'' scenario with no concern to  ethics restrictions. The scenario proposed can either be based on experimental data or defined as an intuition of what might happen in reality. Furthermore, there are also cases where actual models do not match real-world experimentation and they need to be further investigated (in a simulation model). In addition, new hypotheses and research questions may be defined together with immunologists as simulation goals. The objectives come from real-world observation. We assume, however, that real-world observation and experimentation has been previously performed by immunologists.

{\bf 2. Describe the system}. In this step, it is necessary to use documents (immunology books and articles, transcripts of interviews with experts, etc.) describing how the immune elements to be simulated work and interact. The description of the system is based on knowledge acquired by theoretical work, real-world observation and laboratory experimentation. Due to the complexity of the elements and processes in the immune system, however, this knowledge is scarce. The immune system is far from being fully fathomable, and the descriptions found in literature are only partial representations and assumptions of what occurs in reality.

{\bf 3. Investigate existing theories and established models.} In order to build a new simulation model, it is common to look at the existing models and investigate their hypotheses, objectives, validation process and limitations. For some cases, these established models have somehow been validated against experimental data. With this practice it is possible to build a new model as an improvement of what has already been established in order to further investigate a certain immune process.

{\bf 4. Use experimental data.} Currently, most simulation models are built based on real-world observation and experimentation. There are some models, however, where there is no data available (for example, when Jerne's network theory was conceptualized~\cite{Jerne:1974}). These models are based purely on theoretical assumptions with the purpose of providing more insights about what happens in the real world. Furthermore, in the field of immunology, the non-existence of data can be due to the lack of understanding of a process, or a difficulty or even impossibility in collecting information with current technology. In other cases, a hypothesis is first formulated requiring experimental data to confirm it. There is, therefore, the need to collect this data. For instance, Foan {\it et al.}~\cite{Foan:2011} implemented a system dynamics simulation of T cell subsets throughout a person's lifetime based on an established mathematical model developed by Balcheva~\cite{Baltcheva:2010}. The authors conclude that further validation of this model is necessary and so a novel data set should be collected as there are arguably more specific markers that could help to gain further insights from the model.

{\bf 5. Build conceptual model.} The conceptual model of a problem is an abstraction intended to contain the principal aspects observed in the real world, considering the necessary level of details~\cite{Robinson:2008}. In this step we formally define the model scope, the objectives previously outlined, the inputs and outputs and the simplifications. The process of creating a conceptual model evolves with decisions regarding the model scope and level of detail~\cite{Robinson:2004}. The acceptance of the conceptual model should be agreed with immunologists. According to Ulgen {\it et al.}~\cite{Ulgen:1994}, ``{\it rigorous validation procedure for the conceptual model is as important as the verification and validation of the model because it saves time and redirects the simulation developers in the right direction before time is wasted in the study}''. Due to the limitations of a immune simulation, it is important to abstract the relevant real-world features and build a simple model. According to Kotiadis and Robinson~\cite{Robinson:2008}, the importance of model abstraction ``{\it relies on the fact that there is no need to model all that is known about the real problem. Simpler models are developed and run faster, they are flexible, require less data and results are easier to be interpreted}''. The nature of the immune problems thus implies that the model should be developed in order to address a few objectives, within a limited scope. The description of the system (and definition of the conceptual model) should therefore focus on the parts of the immune system (scope, elements, information available, assumptions, hypotheses) relevant to achieve the simulation goals. Daigle discusses the challenges of modelling immunology~\cite{Daigle:2006}. As it is  a field in which information is still being gathered, simulations have to be updated frequently to suit new findings. Moreover, current computational resources and modelling techniques are in development.  It is still thereby impossible to  represent computationally an entire pool of cells of a typical immune response (around $10^{12}$ cells). In addition, immunological systems are mostly hierarchical, involving several layers and complex interactions between the elements of these layers.

{\bf 6. Identify elements, parameters, aggregates, etc. already established in theory and real-world data.} The study of the conceptual model provides a means to understand the problem and the best way to represent the elements of the system. This stage influences the choice of the most appropriate simulation approach. For example, if the in the conceptual model it is established the interactions of the simulation will occur at a cellular population level rather than an individual cell level, this might indicate that a top-down simulation approach would be more suitable to build the model.

{\bf 7. Decide on the most appropriate simulation approach.} This decision is made based on the characteristics of the problems, the research questions to be addressed, the scope, the level of aggregation and the experimental data available. Some of the most common approaches used in immunology are agent-based modelling and simulation, discrete-event modelling and simulation, cellular automata and system dynamics. Cellular automata is used for problems involving autonomous individual interactions within a neighbourhood placed in a lattice and emergent behaviour. Agent-based simulation is suitable for problems involving autonomous individual behaviour, elements spacial localization, memory and emergence. Discrete-event simulation tackles problems that are process-oriented, which have passive individual entities and chronological sequence of events. Furthermore, each event occurs at an instant in time and marks a change of state in the system. It can be used for any experiment where there is no need for continuous time. SD defines a system at a high level of aggregation and, therefore, it should be used when the research question involves patterns of behaviours and feed-back interactions between the aggregates. This approach is very useful to simulate dynamics of populations and interactions between different populations overtime. For example, interactions between tumours and populations of effector cells, populations of viruses and T cells, etc.

{\bf 8. Represent elements, parameters, etc. using the appropriate simulation approach.} Once the simulation approach is chosen, the elements defined in the conceptual model need to be translated into their correspondent implementation used by each approach, for instance, stocks, flows, parameters and information for SD or agents and rules for ABMS. This step is part of the construction of the simulation model, defined in the next step.

{\bf 9. Build the simulation model.} This stage includes the development of the computational implementation of the model in a simulation tool. The implementation is a software representation of the requisites defined in the conceptual model. The computational model is the final product to be used by the immunologists.

{\bf 10. Verify the model.} The model verification is the process of ensuring that the model design has been transformed into a computer model with sufficient accuracy~\cite{Robinson:2004}.

{\bf 11. Validate the model with existing theories and, if available, real-world data.} Validation ensures that the model is sufficiently accurate for the purpose at hand. For immunology it is acknowledged that models are not intended to be completely accurate for a number of reasons: (1) there is no real world data to compare against, (2) there is little data, (3) real-world data is inaccurate, (4) even if the data is accurate, the real world data is only a sample, which in itself creates inaccuracy. Verification and validation are continuous and iterative processes performed throughout the life cycle of a simulation study~\cite{Robinson:2004}.

{\bf 12. Experimental design.} The experimental design improves the experimentation process by the definition of sound experiments with trustworthy results. In this stage, the experimental factors that are most likely to lead to significant improvements are identified. This process is developed using data analysis, expert knowledge, preliminary experimentation and sensitivity analysis.

{\bf 13. Experimentation.} Experimentation is conducted following the experimental design guidelines. It can make use of multiple simulation replications; interactive experimentation (observing the simulation and making changes to the model to see the effects); batch experimentation (setting experimental factors and leaving the models to run for a pre-defined run length); comparing alternatives (where there is a limited number of scenarios to be compared) and search experimentation (when there is no predefined number of scenarios).

{\bf 14. Result Analysis.} Plots and statistics are collected during the simulation. The result analysis is the process that interprets results and the best way to present them.

{\bf 15. Report Findings.} After results are interpreted, there is the need to report the findings from the simulations. For immunology, it can be new insights, verification of a theory, etc.

{\bf 16. Validate and add more requisites with immunologists.} Building an immune simulation is an iterative process. Generally the model is built together with immunologists, and, in every step of the framework, the model elements should be verified with them.

The process of simulation is iterative, as shown in Figure 1. During the model development, additional data might become available, which changes the system description/objectives and impacts on every step of the process. Moreover, as validation occurs throughout the whole process, if any of the stages is not validated (data available, real world understanding and description, conceptual model, computer model, experimental design, etc), there is the need to go back and rethink the invalid state, which impacts on the subsequent steps.

\begin{figure}[!h]
  \vspace{-0.2cm}
  \centering
   {\epsfig{file = 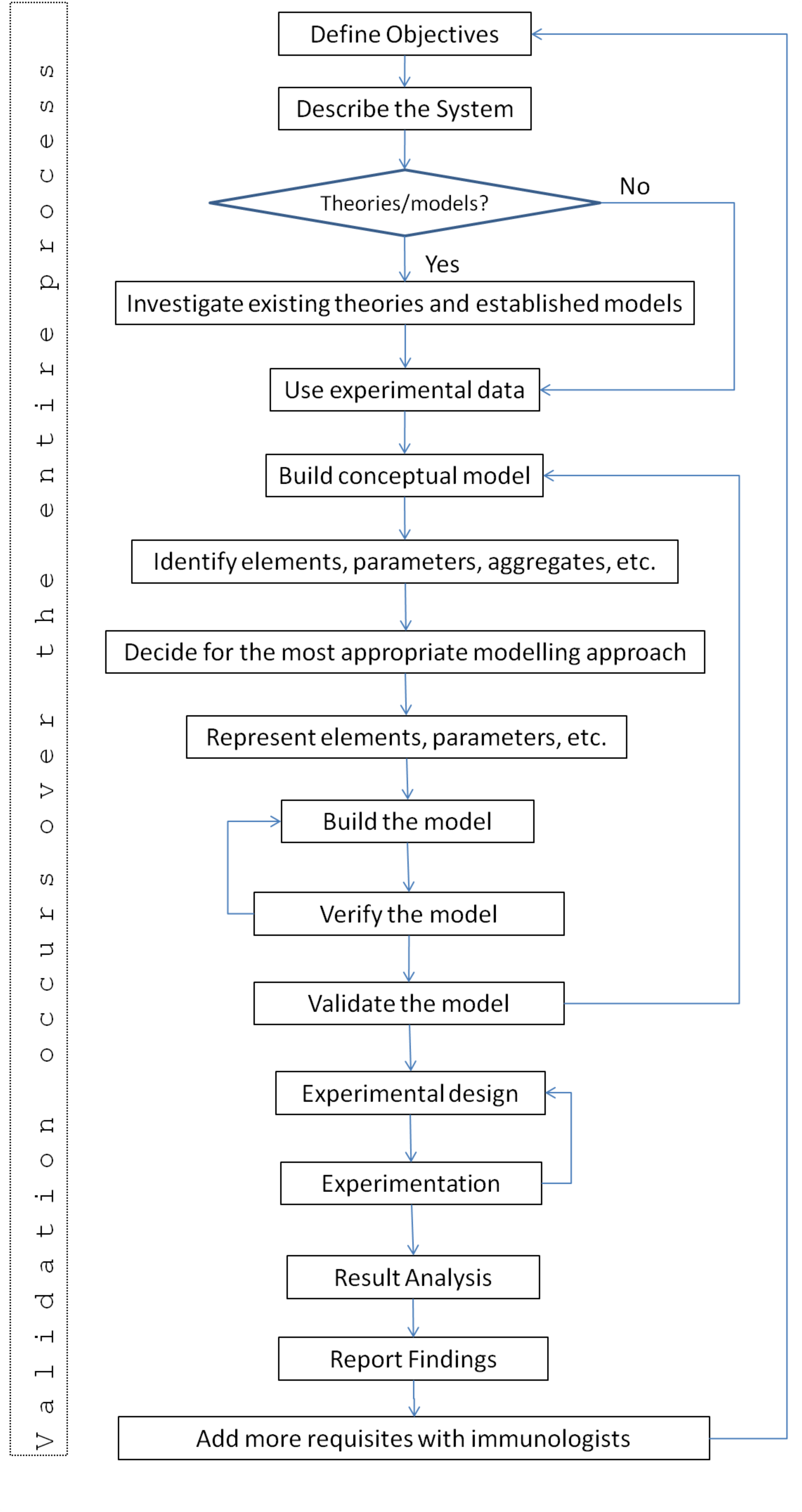, width = 6.5cm}}
   \vspace{-20pt}
  \caption{Process of simulation study: the life cycle.}
  \vspace{-10pt}
  \label{fig:steps}
 \end{figure}

\section{Case Study}
\label{CaseStudy}

The objective of our case study is to exemplify how the steps defined previously can be employed to help in the development of a simulation for the immune system. The problem to be investigated  is how the population of peripheral naive T cells evolve during the course of an individual's lifetime. The development of the simulation model is depicted below.

{\bf 1. Define Objectives.} The simulation goal is concerned with establishing an understanding of the peripheral naive T cell repertoire dynamics over time.

{\bf 2. Describe the System.} In the human body there is a type of white blood cell, namely the naive T cell, which plays an important role in the immune system by responding to new infections in the organism. Before an individual reaches the age of 20, the set of naive T cells is sustained primarily from thymic output. In middle age, however, there is a change in the source of naive T cells: as the thymus involutes, there is a considerable shrinkage in its T cell output, which means that new T cells are mostly produced by peripheral expansion. There is also a belief that some memory T cells have their phenotype reverted back to the naive proliferation cells type~\cite{Murray:2003}. Furthermore, memory cells are originated from active T cells.

{\bf 3. Investigate existing theories and models.} Immunologists found out that thymic contribution in an individual are quantified by the level of a biological marker called `T cell receptors excision circle' (TREC). TREC is circular DNA originated during the formation of the T-cell receptor. The percentage of T cells possessing TRECs decays with shrinkage of thymic output, activation and reproduction of naive T cells~\cite{Murray:2003}. This means that naive T cells originating from the thymus have a greater percentage of TREC than those originating through other proliferation and with time there is a depletion of naive T cells from thymus in the organism. There is an existing model proposed by Murray~\cite{Murray:2003} that investigates the thymic output and decay of these cells mathematically with the use of an ODE system.

{\bf 4. Experimental data.} TREC data collected by immunologists is presented by Murray {\it et al.}~\cite{Murray:2003}, which also develops an ODE system model for the dynamics of peripheral naive T cells. Furthermore, the authors provided us with data on active cells and total naive T cells in individuals with age ranging from $1$ to $55$ years. If we assume that this data has been validated by the immunologists and expresses what occurs in reality, we can use this information for the continuation of our investigations on naive cells from peripheral proliferation.

{\bf 5. Conceptual model.} The population dynamics of the model is shown in Figure 2. We have four main populations of cells: naive from thymus (in the figure naive), naive from proliferation, active cells and memory cells. Naive and memory cells are sources of naive cells from proliferation. Active cells are sources of memory cells. The scope of the simulation, therefore, is limited to the dynamics of these four populations. The number of naive from thymus and active cells are given by real-world data collected by immunologists. The conceptual model and data used are the same as the those from the existing model proposed by Murray~\cite{Murray:2003}. The objective is to determine what are the dynamics of the naive cells from peripheral proliferation with age under a systems simulation approach different from the ODEs simulation. Another goal is to determine the rates in which the naive cells from thymus and memory cells become naive cells from proliferation; and the rates in which active cells become memory cells.

\begin{figure}[!htpb]
\vspace{-20pt}
 \begin{center}
  \resizebox{5cm}{!}{\includegraphics{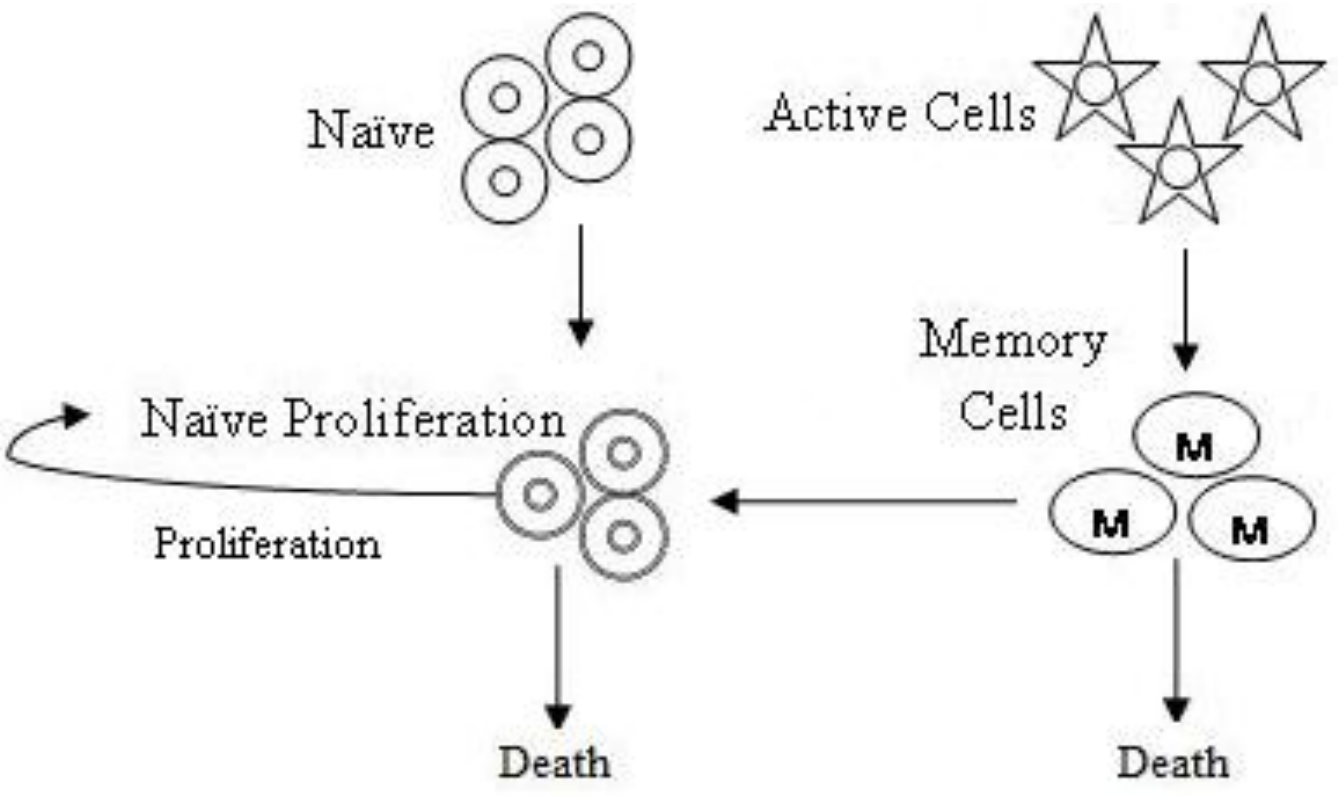}}
 \end{center}
 \label{fig:Conceptual}
 \vspace{-20pt}
 \caption{Dynamics of Naive T cells.}
 \vspace{-20pt}
\end{figure}

{\bf 6 and 7. Identify elements, parameters, etc. and decide on the most appropriate simulation approach.} As the investigations regard populations dynamics at a high level of aggregation, as defined in the conceptual model, we decided to build the simulation using the system dynamics approach, where the aggregates will be each different cell population and the feedback loops are those represented by the arrows in the conceptual model of figure 2.

{\bf 8. Represent elements, parameters, etc. using the appropriate simulation approach.} As stated before, system dynamics models consist of stock, flows, information and feedback loops. The stocks of the simulation are the variables that accumulate over time; flows modify the stocks by adding or subtracting elements. For the simulation, therefore, the stocks will be the number of naive cells from proliferation and memory cells. The stock of naive cells from proliferation is modified by flows such as addition of new cells from thymus, peripheral proliferation and death, as shown in Figure 2 from the conceptual model. The memory cells stock is changed by the flows: active cells reversing into memory and memory cells death. Both naive and active cells are represented in the simulation as look up tables. Moreover, parameters representing rates in which flows modify the stocks need to be incorporated. These parameters are: $NaiveThymusProliferationRate$, $NaiveProliferationRate$, $NaiveProliferationDeathRate$, $MemoryToNPRate$, $MemoryDeathRate$ and $ReversionToMemoryRate$.

{\bf 9. Build the simulation model.} The final SD model implemented is shown in Figure 3.

  \begin{figure*}[!htpb]
    \vspace{-20pt}
    \centering
    \resizebox{9cm}{!}{\includegraphics{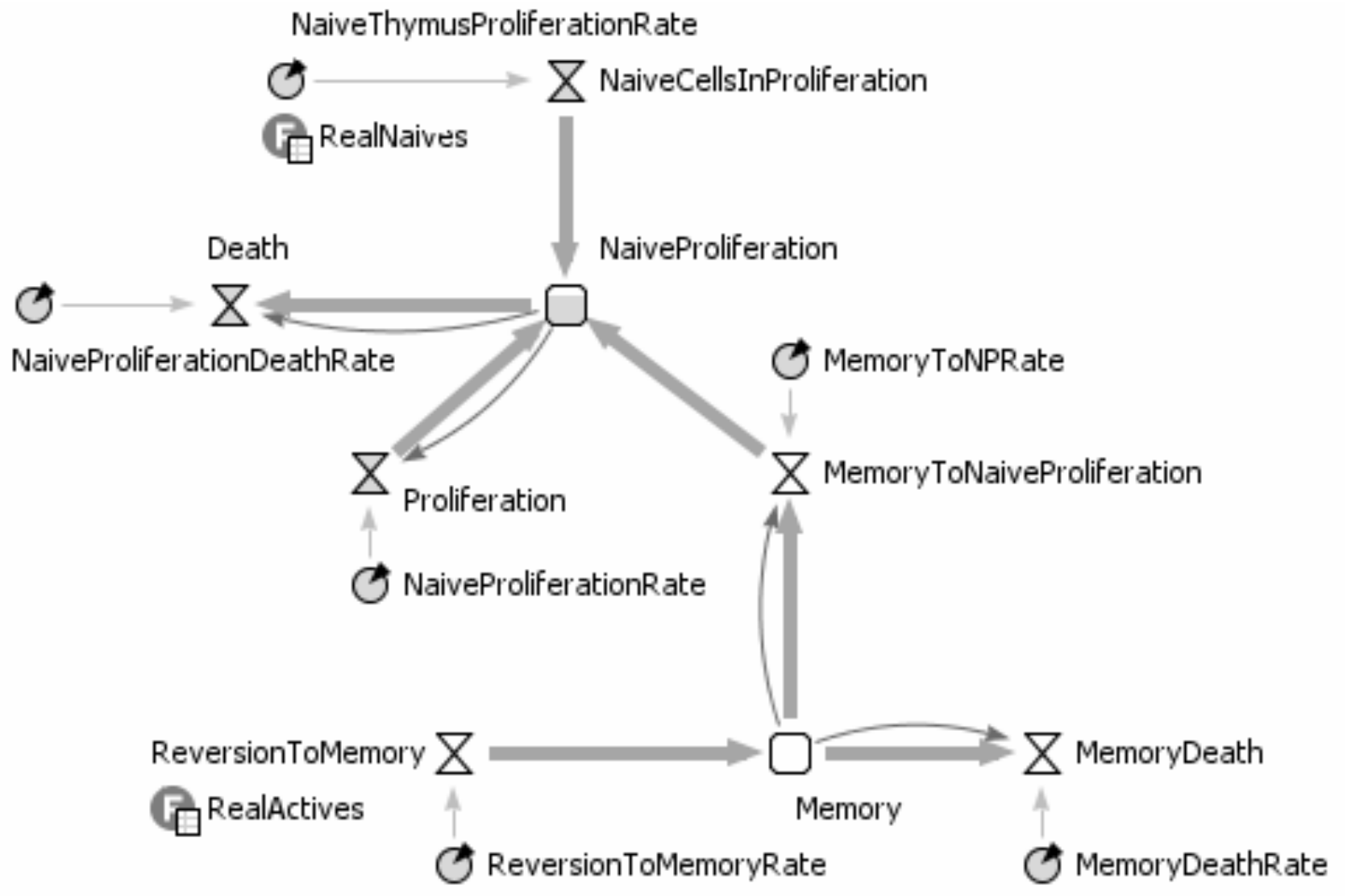}}
    \vspace{-10pt}
    \caption{SD built from naive T cells data.}
    \vspace{-20pt}
    \label{fig:SDNaiveFromScratch}
  \end{figure*}

In the figure, the stocks are represented by the box~\includegraphics[scale=0.7]{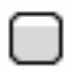}, the flow variables are represented by the hourglass~\includegraphics[scale=0.7]{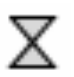}, flow~\includegraphics[scale=0.7]{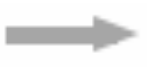},  parameters~\includegraphics[scale=0.7]{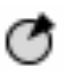} and information~\includegraphics[scale=0.7]{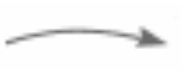}. Information indicates that the stock value is used in the flow calculation. RealNaives and RealActives are look-up tables containing the experimental data.

{\bf 9. Verify the model.} As the model studied is quite simple, we verified our implementation against the conceptual model. Further verification was performed during the experimental stage.

{\bf 10. Validate the model with existing theories and, if available, real-world data.} As we mentioned, the validation process is performed throughout all simulation development. We validate our results against the data set provided for the total number of naive T cells in the organism, as shown in the result analysis (step 12). 

{\bf 11. Experimental design and experimentation.} For this case study we will run one experiment in which we adjust the parameters to fit the original data. The simulation was run for a period correspondent to sixty years.

{\bf 12. Result Analysis and findings.} We calibrated the simulation parameters against the data provided and the results obtained are shown in Figure 4(a). Figure 4(b) shows the results obtained in~\cite{Murray:2003}. For both approaches, the total number of naive cells for the simulation are different for the first eighteen years and this difference as well as the parameters obtained need further investigation with immunologists. After twenty years, the numbers for the real data observed and the outcomes obtained are very close. Moreover, it is possible to observe how the population of naive cells from proliferation grows with time and is prevalent in the total naive cells cohort after fifty years.

\begin{figure}[!htpb]
\vspace{-20pt}
 \begin{center}
\subfigure[]{
   \includegraphics[scale =0.38] {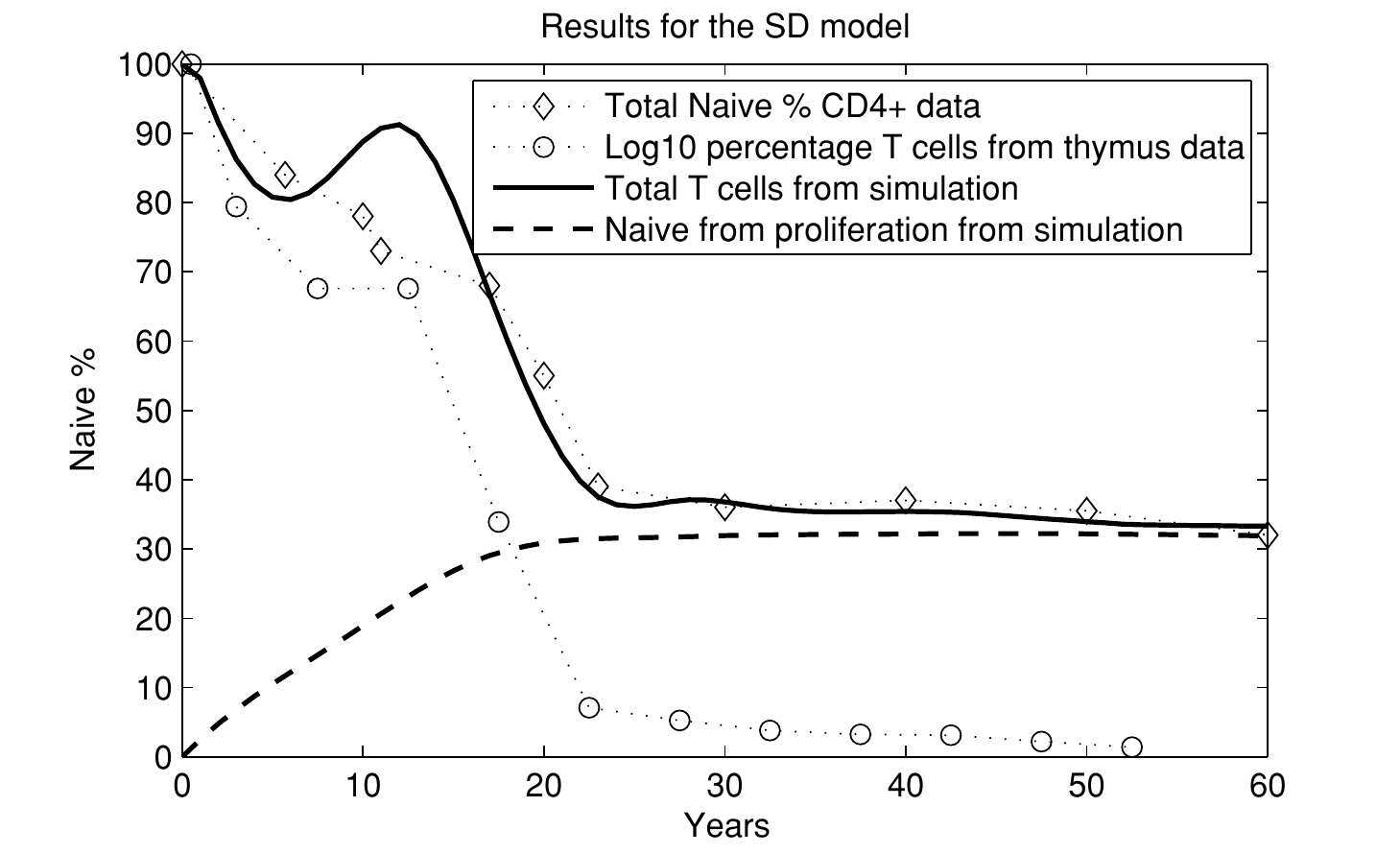}
 }
 \subfigure[]{
   \includegraphics[scale =0.38] {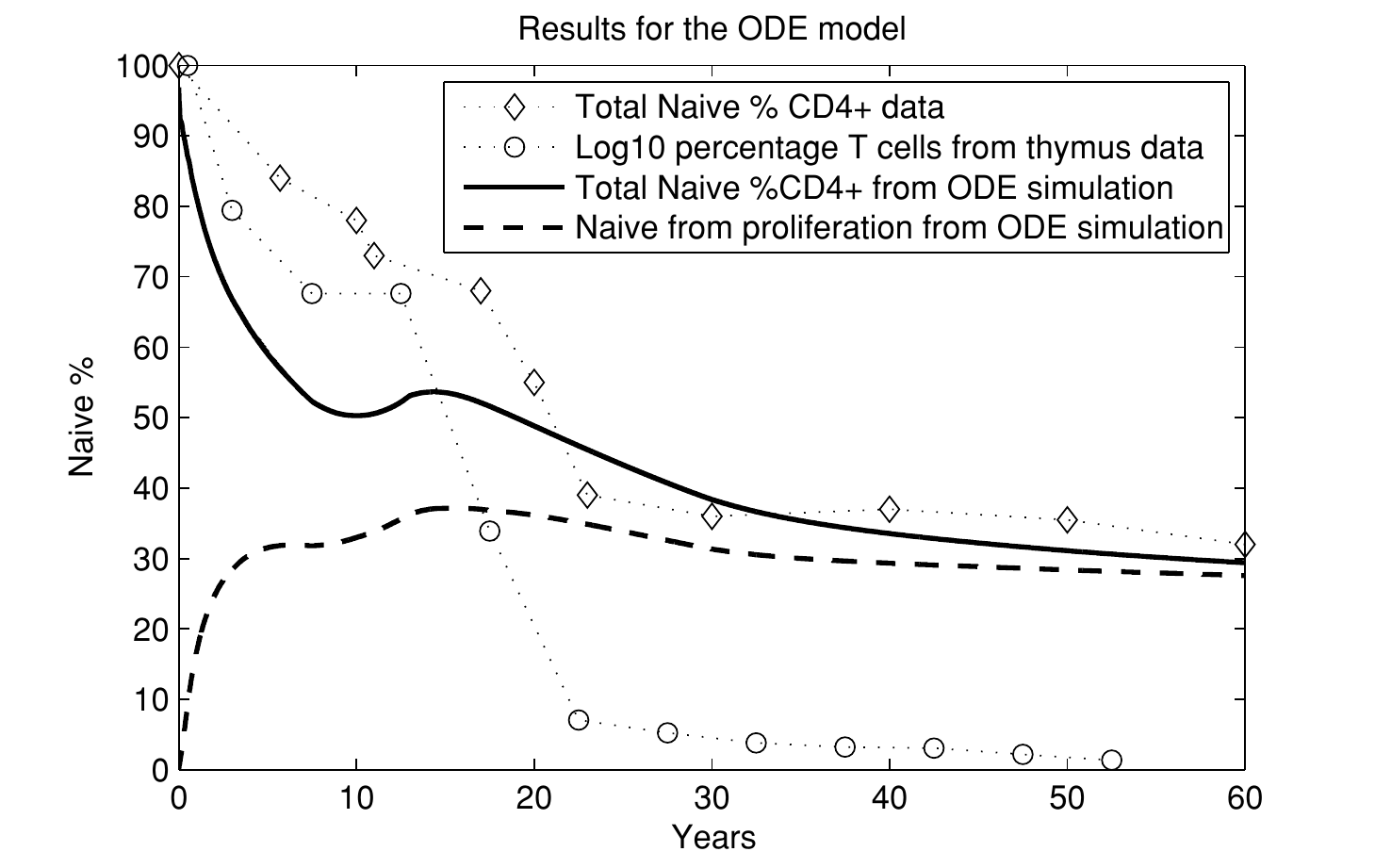}
 }
 \end{center}
 \label{fig:Conceptual}
 \vspace{-25pt}
 \caption{{\small Simulation results. For the SD simulation, the parameter values used are: NaiveThymusProliferationRate = 0.025, NaiveProliferationDeathRate = 0.017, MemoryToNPRate = 0.001, ReversionToMemoryRate = 0 and MemoryDeathRate = 0.05.}}
 \vspace{-20pt}
\end{figure}

{\bf 13. Validate and add more requisites with immunologists.} Results (and parameters calibration) validation needs to be done with immunologists, as well as new requisites do be added in the simulation model. There is the need to investigate whether the simulation models are informative and relatively accurate. Further research also needs to be done to explain the outcome differences for both models.

\section{Conclusions}
\label{Conclusions}

Although there are examples showing the success of simulation aiding advances in immunology, this set of methodologies is not popular among immunologists. The overall objective of our research, therefore, is to outline the potential contribution of simulation methods to help immunological studies and invite experts (simulation developers, computer scientists, etc) to build solutions in this field. For this paper, we had two research goals. The first goal was to show that there is a distinct place for simulation in the tool set used by immunologists and present the most common simulation approaches. As there are no general guidelines for the development of immune simulations, our second objective was to introduce our own guidances for conducting simulation studies in immunology and outline the pitfalls that might be encountered during the development of a simulation model. We achieved our first objective by arguing that popular methodologies used such as ODEs and reductionist methods have limitations that are overcome by simulation. Moreover, we argued that for many simulation methods, the problem representation is closer to the systems natural description. For our second goal, we studied several simulations using different approaches in the literature and outlined their common features. This helped us to develop a simulation life-cycle with several steps to be followed. These steps encompass common aspects to be considered during the development of immune simulations, independent of the simulation approach adopted. As future work we want to improve the life-cycle introduced and develop a decision framework that further helps with the choice of a simulation approach according to the problem presented. In addition, we aim at making our set of guidances more specific according to the simulation method used.

\bibliographystyle{splncs}
\bibliography{tese}

\end{document}